\documentstyle[aps,preprint,tighten]{revtex}

\begin{document}

\draft

\title{Exact Green's functions for delta-function potentials and 
renormalization in quantum mechanics}
\author{R. M. Cavalcanti\footnote{Present address: Instituto de
F\'\i sica, Universidade de S\~ao Paulo, Cx. Postal 66318,
CEP 05315-970, S\~ao Paulo, SP. E-mail: rmoritz@fma.if.usp.br}}
\address{Institute for Theoretical Physics,
University of California, Santa Barbara, CA 93106-4030, USA}
\maketitle

\begin{abstract}
We present a simple recipe to construct the Green's function
associated with a Hamiltonian of the form $H=H_0+\lambda\,\delta(x)$,
where $H_0$ is a Hamiltonian for which the associated Green's
function is known. We apply this result to the case in which $H_0$ is
the Hamiltonian of a free particle in $D$ dimensions. Field theoretic 
concepts such as regularization, renormalization,
dimensional transmutation and triviality are introduced naturally in 
order to deal with an infinity which shows up in the formal
expression of the Green's function for $D\ge 2$.
\end{abstract}




\section{Introduction}

Quantum electrodynamics (QED), like most quantum field theories,
is plagued with infinities. It was a major accomplishment when
Schwinger, Feynman, Tomonaga and Dyson \cite{Schweber} showed how 
to extract meaningful information from QED, in spite of such
infinities.
In a {\em renormalizable} theory, like QED, such infinities can be
dealt with in a two-step process: (i) its short distance
(or high energy) behavior is modified with the introduction of
a cut-off, which gives rise to finite answers, and (ii)
the parameters of the theory are redefined in order to
absorb the divergences which appear when the cut-off is
removed. These steps are the {\em regularization} and
{\em renormalization} of the theory, respectively.
Infinities of this sort also occur in non-relativistic quantum
mechanics
if the potential is singular enough, for instance the
Dirac delta-function potential in two or more dimensions 
\cite{Albeverio,Jackiw,Tarrach,Perez,Mead,Manuel,Adhikari,Park}
and the Aharonov-Bohm potential \cite{Manuel,Park}.
This provides a unique framework in which the important
concepts of regularization and renormalization
can be explained free 
from the technical complications
usually found in quantum field theory.

In this paper we study the Dirac delta-function potential.
This problem has been previously studied in the literature using 
a variety of techniques: exact solutions of
the Schr\"odinger equation 
\cite{Tarrach,Perez} (or its integral version, the Lippman-Schwinger 
equation \cite{Mead,Manuel,Adhikari}), the self-adjoint extension 
method \cite{Albeverio,Jackiw}, and
Green's function techniques \cite{Park,Cavalcanti}.
Here we use the latter, which have a closer resemblance with the 
techniques usually employed in quantum field theory.
Besides that, it is very easy
to find the Green's function associated with a Hamiltonian
of the form $H=H_0+\lambda\,\delta({\bf x})$ when the Green's
function associated with $H_0$ is known; this is the content of
Sec.~II. In Sec.~III, we apply
this result to the case in which $H_0$ is the Hamiltonian
of a free particle in $D$ dimensions. For $D\ge 2$ an infinity
shows up in the formal expression of the Green's function.
In the two- and three-dimensional cases,
this infinity can be removed in a consistent way,
in what amounts to be a simple exercise in
regularization and renormalization.
Finally, in the Appendix we show how the technique presented in
Sec.~II can be adapted to the calculation of the scattering amplitude.


\section{Green's functions for delta-function potentials}

The Green's function $G(E;{\bf x},{\bf y})$ associated with the 
Hamiltonian $H$ is the solution of the differential equation
\begin{equation}
\label{Gdef}
(E-H)\,G(E;{\bf x},{\bf y})=\delta({\bf x}-{\bf y})
\end{equation}
satisfying the boundary condition 
\begin{equation}
\lim_{|{\bf x}-{\bf y}|\to\infty}\,G(E;{\bf x},{\bf y})=0.
\end{equation}
Here {\bf x} and {\bf y} are points in $D$-dimensional euclidean space
and,
correspondingly, $\delta({\bf x}-{\bf y})$ is a $D$-dimensional 
Dirac delta-function.

One can use the completeness of the eigenfunctions of $H$
to write the solution of (\ref{Gdef}) as
\begin{equation}
\label{Gsol}
G(E;{\bf x},{\bf y})=\sum_n\frac{\psi_n({\bf x})\,
\psi_n^*({\bf y})}{E-E_n},
\end{equation}
where $E_n$ and $\psi_n$ are the eigenvalues and eigenfunctions
of $H$, respectively. Now,
let us suppose that the Hamiltonian can be written as the sum
of two terms,
\begin{equation}
\label{H}
H=H_0+\lambda\,\delta({\bf x}),
\end{equation}
and that the Green's function $G_0(E;{\bf x},{\bf y})$ associated with
$H_0$ is known. Then, as we shall show below, there is a
very simple recipe to write $G$ in terms of $G_0$.

Let us rewrite Eq.~(\ref{Gdef}) in integral form
\cite{Economou}
(for simplicity we omit the dependence on $E$):
\begin{eqnarray}
\label{quasiG}
G({\bf x},{\bf y})&=&G_0({\bf x},{\bf y})
+\int d^Dz\,G_0({\bf x},{\bf z})\,\lambda\,
\delta({\bf z})\,G({\bf z},{\bf y})
\nonumber 
\\
&=&G_0({\bf x},{\bf y})+\lambda\,G_0({\bf x},{\bf 0})\,
G({\bf 0},{\bf y}).
\end{eqnarray}
Now we put ${\bf x}={\bf 0}$ in the expression above and 
solve it for $G({\bf 0},{\bf y})$; then, inserting the result 
in (\ref{quasiG}), we obtain an
explicit expression for the Green's function associated with $H$:
\begin{equation}
\label{G}
G({\bf x},{\bf y})=G_0({\bf x},{\bf y})
+\frac{G_0({\bf x},{\bf 0})\,G_0({\bf 0},{\bf y})}
{\frac{1}{\lambda}-G_0({\bf 0},{\bf 0})}.
\end{equation}
It is worth to note that successive applications of this
procedure allows one to find the Green's function for a potential
with an arbitrary number of delta-functions.


\section{Bound states and renormalization}

In this section, we shall investigate the bound states of 
the Hamiltonian (\ref{H}),
with $H_0$ the Hamiltonian of a free particle in $D$ dimensions
(we use units such that $\hbar=2m=1$):
\begin{equation}
H_0=-\nabla^2\equiv-\sum_{j=1}^{D}\frac{\partial^2}{\partial x_j^2}.
\end{equation}

It follows from (\ref{Gsol}) that the energy levels of bound 
states are given by the real poles of the Green's function. 
Since there are no bound states in the free particle problem, 
such poles can only appear as zeros of the denominator of the 
second term on the
r.h.s.~of Eq.~(\ref{G}). In order to obtain $G_0(E;{\bf x},{\bf y})$,
we
Fourier transform Eq.~(\ref{Gdef}) (with $H$ replaced with $H_0$),
thus finding
\begin{equation}
\label{G0}
G_0(E;{\bf x},{\bf y})=\int\frac{d^Dk}{(2\pi)^D}\,
\frac{e^{i{\bf k}\cdot({\bf x}-{\bf y})}}{E-k^2}.
\end{equation}
Therefore, in order to find the energy of the bound states we must
solve the equation ($K^2\equiv-E$)
\begin{equation}
\label{poles}
\frac{1}{\lambda}+\int\frac{d^Dk}{(2\pi)^D}\,\frac{1}{k^2+K^2}=0.
\end{equation}
In what follows, we shall examine Eq.~(\ref{poles}) for different
values of $D$.


\subsection{$D=1$}

Performing the integral in Eq.~(\ref{poles}), we find
\begin{equation}
\label{int1D}
\int_{-\infty}^{\infty}\frac{dk}{2\pi}\,\frac{1}{k^2+K^2}=\frac{1}{2K}.
\end{equation}
Then, solving Eq.~(\ref{poles}) for $K$, we obtain 
\begin{equation}
K=-\frac{\lambda}{2}\quad{\rm and}\quad E_B=-\frac{\lambda^2}{4}.
\end{equation}
Note that $\lambda$ must be negative, for $K$ was implicitly
taken positive in Eq.~(\ref{int1D}). Physically, this means
that the potential must be attractive in order to create a bound
state.

For the sake of comparison, we present an alternative,
more elementary derivation of this result. The time-independent
Schr\"odinger equation for a particle in the potential
$V(x)=\lambda\,\delta(x)$ is
\begin{equation}
\label{Sch}
-\frac{d^2}{dx^2}\,\psi(x)+\lambda\,\delta(x)\,\psi(x)=E\,\psi(x).
\end{equation}
For $x\ne 0$ this equation is that of a free particle; 
solving
it for $E=-K^2<0$ and imposing continuity at the origin, we
find
\begin{equation}
\psi(x)=A\,e^{-K|x|}.
\end{equation}
A restriction on the possible values of $K$ is obtained by
integrating Eq.~(\ref{Sch}) from $-\varepsilon$ to $+\varepsilon$
and letting $\varepsilon\to 0^+$:
\begin{equation}
-\psi'(0^+)+\psi'(0^-)+\lambda\,\psi(0)=0,
\end{equation}
from which follows that $K=-\lambda/2$. Therefore, both
methods give the same results.


\subsection{$D=2$}

In this case, a problem occurs that is absent in $D=1$:
$G_0(E;{\bf 0},{\bf 0})$ is (logarithmically) divergent. To deal with
this
problem, we must introduce a cut-off in the integral which
appears in Eq.~(\ref{poles}) and absorb the dependence
on the cutoff in a redefinition of the parameters of the theory
(in this case, the ``coupling constant'' $\lambda$). In
quantum field theory this procedure is known as {\em regularization}
and {\em renormalization}. Let us demonstrate it explicitly.
The first step is to regularize the integral:
\begin{eqnarray}
\int\frac{d^2k}{(2\pi)^2}\,\frac{1}{k^2+K^2}
&=&\frac{1}{2\pi}\,\int_{0}^{\Lambda}\frac{k\,dk}{k^2+K^2}
\nonumber 
\\
&=&\frac{1}{4\pi}\,\ln\left(\frac{\Lambda^2+K^2}{K^2}\right).
\end{eqnarray}
The next step is to absorb the divergent part of the above result
in a redefinition of the coupling constant:
\begin{equation}
\label{ren2}
\frac{1}{\lambda_R}\equiv\frac{1}{\lambda}
+\frac{1}{4\pi}\,\ln\left(\frac{\Lambda^2}{\mu^2}\right).
\end{equation}
(The parameter $\mu$ is arbitrary, and is introduced in order to
keep the argument of the logarithm dimensionless.)
We now take the limit $\Lambda\to\infty$, varying the {\em bare}
coupling constant $\lambda$ in such away that the {\em renormalized} 
coupling constant $\lambda_R$ remains finite; 
Eq.~(\ref{poles}) then becomes
\begin{equation}
\frac{1}{\lambda_R}-\frac{1}{4\pi}\,\ln\left(\frac{K^2}{\mu^2}\right)=0.
\end{equation}
Solving this equation for $K^2$ we find the energy of the bound state:
\begin{equation}
\label{EB}E_B=-K^2=-\mu^2\,\exp\left(\frac{4\pi}{\lambda_R}\right).
\end{equation}

A curious thing happens here: although the Hamiltonian contains
only one parameter ($\lambda$), we have obtained an energy
($E_B$) depending on {\em two} parameters ($\lambda_R$ and $\mu$). 
However, this doubling of parameters is illusory. In fact, it is 
possible to show that the Green's function depends on a sole parameter
(besides $E$, {\bf x} and {\bf y}, of course).
To see this, let us write the denominator of the second term
on the r.h.s.~of Eq.~(\ref{G}) in regularized form:
\begin{eqnarray}
\frac{1}{\lambda}-G_0(E;{\bf 0},{\bf 0})
&=&\frac{1}{\lambda}+\frac{1}{2\pi}\,
\int_0^{\Lambda}\frac{k\,dk}{k^2-E}
\nonumber 
\\
&=&\frac{1}{\lambda}+\frac{1}{4\pi}\,
\ln\left(\frac{\Lambda^2-E}{-E}\right).
\end{eqnarray}
On the other hand, according to Eqs.~(\ref{ren2})--(\ref{EB}),
$\lambda^{-1}=-(1/4\pi)\,\ln(-\Lambda^2/E_B)$; substituting
this in the expression above and
taking the limit $\Lambda\to\infty$, we obtain
\begin{equation}
\frac{1}{\lambda}-G_0(E;{\bf 0},{\bf 0})
=-\frac{1}{4\pi}\,\ln\left(\frac{E}{E_B}\right).
\end{equation}

This is an instance of the so-called {\em dimensional transmutation\/}
\cite{Coleman}:
having started with a Hamiltonian containing only dimensionless
parameters (in this case, the coupling constant $\lambda$), 
we ended up with a theory
containing a dimensionful parameter ($E_B$). 
This happens because during the renormalization process
we had to introduce the dimensionful parameter $\mu$,
thus breaking the scale invariance of the theory.


\subsection{$D=3$}

As in the case $D=2$, $G_0(E;{\bf 0},{\bf 0})$ is divergent, 
and so the bare Green's function, given by Eq.~(\ref{G}), is
ill-defined.
To deal with this problem, we proceed as in the previous
subsection: we regularize the integral which appears
in Eq.~(\ref{poles}),
\begin{eqnarray}
\int\frac{d^3k}{(2\pi)^3}\,\frac{1}{k^2+K^2}
&=&\frac{1}{2\pi^2}\,\int_{0}^{\Lambda}\frac{k^2\,dk}{k^2+K^2}
\nonumber 
\\
&=&\frac{1}{2\pi^2}\left[\Lambda-K\,
\arctan\left(\frac{\Lambda}{K}\right)\right],
\end{eqnarray}
and absorb the divergent part of this result in a redefinition of the
coupling constant,
\begin{equation}
\label{ren3}
\frac{1}{\lambda_R}\equiv\frac{1}{\lambda}+\frac{\Lambda}{2\pi^2}.
\end{equation}
Taking the limit $\Lambda\to\infty$ in Eq.~(\ref{poles})
while keeping $\lambda_R$ fixed, we obtain
\begin{equation}
\frac{1}{\lambda_R}-\frac{K}{4\pi}=0,\end{equation}
from which it follows that
\begin{equation}
K=\frac{4\pi}{\lambda_R}\quad{\rm and}\quad
E_B=-\left(\frac{4\pi}{\lambda_R}\right)^2.
\end{equation}

Here too we can eliminate all reference to $\lambda_R$ in
favor of $E_B$, the energy of the bound state. In fact,
as a simple calculation shows, the denominator of the second
term on the r.h.s.~of Eq.~(\ref{G}) can be rewritten as
\begin{equation}
\label{den3}
\frac{1}{\lambda}-G_0(E;{\bf 0},{\bf 0})
=\frac{\sqrt{-E_B}-\sqrt{-E}}{4\pi}.
\end{equation}


\subsection{$D\ge 4$}

Let us now consider $D=4$. In this case, the regularized form of
the integral in Eq.~(\ref{poles}) is
\begin{equation}
\frac{1}{8\pi^2}\,\int_{0}^{\Lambda}\frac{k^3\,dk}{k^2+K^2}
=\frac{1}{16\pi^2}\left[\Lambda^2-K^2\,
\ln\left(\frac{\Lambda^2+K^2}{K^2}\right)\right].
\end{equation}
The quadratically divergent term $\Lambda^2/16\pi^2$ may be
absorbed in a redefinition of the coupling constant,
similar to (\ref{ren2}) and (\ref{ren3}). However,
the second term on the r.h.s., which is also
divergent, cannot be eliminated this way, as it is intrinsically
dependent on $K$. The same problem occurs for $D>4$.
This is part of Friedman's theorem \cite{Friedman}:
it is not possible to define a contact (i.e., zero-range) potential 
in more than three dimensions possessing bound states with finite
energy. 
To present the other piece of that theorem,
let us note that in order that $\lambda_R$ be finite (in $D=2$ and 3),
$\lambda$ must tend to $0^-$ when the cut-off is removed 
[see Eqs.~(\ref{ren2})
and (\ref{ren3})]. On the other hand, if one insists to keep 
$\lambda$
finite, then one of these two alternatives will follow: 
(i) if $\lambda<0$,
the Hamiltonian is unbounded from below (the energy of the bound state
depends on the cut-off $\Lambda$, and tends to $-\infty$ when 
$\Lambda\to\infty$), or (ii) if $\lambda>0$, there is no way to avoid
the denominator of the second term on the r.h.s.~of Eq.~(\ref{G}) of
diverging, and so the Green's function
is the same as that in the absence of the 
potential.\footnote{By treating the delta-function as the limit of
a sequence of spherical cores, one can verify \cite{Huang} that in 
two or more dimensions
the $s$-waves are the same as those of a free particle, except
at the origin, where they drop to zero discontinuously (the other
partial waves are not affected by the potential, since
they vanish at the origin). Therefore, the
assertion made in Ref.\cite{Tarrach}, that a repulsive delta-function 
potential in two or more dimensions expels the $s$-waves from the 
Hilbert space, is wrong.}
This is precisely the other piece of Friedman's theorem:
a repulsive delta-function potential in more than one dimension
does not scatter, and so it is said to be {\em trivial\/}.


\acknowledgments

I thank Carlos Alberto Arag\~ao de Carvalho
for a critical reading of this paper. This work had financial 
support from the Conselho Nacional de Desenvolvimento Cient\'\i fico e
Tecnol\'ogico, and was supported in part by the National Science
Foundation under Grant No.~PHY94-07194.


\appendix

\section{}

Here we show how to obtain scattering amplitudes
for a Dirac delta-function potential. First let us
recall that for positive energies $E=k^2$ the Schr\"odinger
equation is equivalent to the Lippman-Schwinger equation
\begin{equation}
\label{LS}
\psi({\bf x})=\psi_0({\bf x})+\int d^Dx'\,G_0^R({\bf x},{\bf x}')\,
V({\bf x}')\,\psi({\bf x}'),
\end{equation}
where $\psi_0$ is a solution of the free Schr\"odinger equation 
and 
$G_0^R({\bf x},{\bf x}')\equiv G_0(E+i\varepsilon;{\bf x},{\bf x}')$ 
is the free retarded Green's function.
For $V({\bf x}')=\lambda\,\delta({\bf x}')$, Eq.~(\ref{LS}) gives
\begin{equation}
\label{psi}
\psi({\bf x})=\psi_0({\bf x})+\lambda\,
G_0^R({\bf x},{\bf 0})\,\psi({\bf 0}).
\end{equation}
Now we put ${\bf x}={\bf 0}$ in this expression and solve it for
$\psi({\bf 0})$; inserting the result in (\ref{psi}), we
obtain
\begin{equation}
\label{psis}
\psi({\bf x})=\psi_0({\bf x})+\frac{G_0^R({\bf x},{\bf 0})\,
\psi_0({\bf 0})}{\frac{1}{\lambda}-G_0^R({\bf 0},{\bf 0})}.
\end{equation}
Note that for two and three dimensions the denominator of this term is
divergent, but it becomes finite after renormalization.

From the asymptotic behavior of this expression as 
$r\equiv|{\bf x}|\to\infty$
one can extract the scattering amplitude. As an example, 
let us consider the three-dimensional scattering problem. 
In this case, the free Green's function is given by
\begin{equation}
G_0(E+i\varepsilon;{\bf x},{\bf x}')
=-\frac{e^{i\sqrt{E}\,|{\bf x}-{\bf x}'|}}{4\pi\,|{\bf x}-{\bf x}'|}.
\end{equation}
Substituting this expression and Eq.~(\ref{den3}) in Eq.~(\ref{psis}),
taking $\psi_0({\bf x})=\exp(ikz)$ (which represents a particle
moving along the positive $z$ direction), and
comparing the result with the asymptotic expression of the
wave function,
\begin{equation}
\psi({\bf x})\approx e^{ikz}+f(k,\theta,\phi)\,\frac{e^{ikr}}{r}
\qquad(r\to\infty),
\end{equation}
we find the following expression for the scattering amplitude:
\begin{equation}
f(k,\theta,\phi)=-\frac{1}{\sqrt{-E_B}-ik}.
\end{equation}
As expected, the scattering is isotropic, since only the $s$-waves
``see'' the zero-range potential located at the origin. 



\end{document}